\documentclass[prb, superscriptaddress,amsmath,amssymb,twocolumn,amsfonts,floatfix, longbibliography]{revtex4-2}
\usepackage{mathtools}
\usepackage{braket}
\usepackage[dvipsnames]{xcolor}
\usepackage{float}
\usepackage{subfigure}
\usepackage{upgreek}
\usepackage{pgffor}
\usepackage{mathrsfs}
\usepackage{float}
\usepackage{silence}
\usepackage[colorlinks=true,linktoc=page,linkcolor=blue,citecolor=magenta,urlcolor=Fuchsia]{hyperref}
\usepackage[normalem]{ulem}
\usepackage{sidecap,tikz}
\usepackage{orcidlink}
\graphicspath{{Old/}}

\newcommand{\beq}{\begin{equation}}
\newcommand{\eeq}{\end{equation}}
\newcommand{\bea}{\begin{eqnarray}}
\newcommand{\eea}{\end{eqnarray}}

\newcommand{\fig}[1]{Fig.~\ref{#1}}
\newcommand{\sect}[1]{Sec.~\ref{#1}}
\newcommand{\tab}[1]{Table~\ref{#1}}

\mathchardef\mhyphen="2D 

\begin{document}


\title{Gyrotropic Fingerprints of Magnetic Topological Insulator–Unconventional Magnet Interfaces}

\author{{Neelanjan Chakraborti}\,\orcidlink{0009-0004-5710-0739}}
\email{neelanjanc23@iitk.ac.in}
\affiliation{Department of Physics, Indian Institute of Technology, Kanpur 208016, India}
\author{{Snehasish Nandy}}
 \thanks{Jointly supervised this work}
 \email{snehasish@phy.nits.ac.in}
\affiliation{Department of Physics, National Institute of Technology Silchar, Assam 788010, India}
\author{{Sudeep Kumar Ghosh}\,\orcidlink{0000-0002-3646-0629}}
 \thanks{Jointly supervised this work}
\email{skghosh@iitk.ac.in}
\affiliation{Department of Physics, Indian Institute of Technology, Kanpur 208016, India}

\begin{abstract}
Unambiguously identifying unconventional magnetic orders requires probes directly sensitive to their momentum-dependent spin-split band structures. Here, we employ a Zeeman quantum geometry framework to study magnetotransport at the interface between a magnetic topological insulator and an unconventional magnetic insulator. By choosing the magnetic layer to be insulating, the transport response originates solely from the proximity-induced exchange field, eliminating contributions from itinerant magnetic carriers. We focus on the linear intrinsic gyrotropic magnetic (IGM) response, which decomposes into conduction and displacement current components governed by the Zeeman Berry curvature and Zeeman quantum metric, respectively. We uncover a universal hierarchy in which the transverse displacement IGM response exhibits characteristic even-fold angular harmonics for magnetic orders ranging from $p$- to $i$-wave, while the longitudinal IGM response distinguishes the parity of the magnetic order through robust sign-reversal patterns. In contrast, the conduction IGM component remains largely insensitive to the underlying magnetic symmetry. Consequently, the displacement IGM current emerges as a high-fidelity symmetry fingerprint of unconventional magnetic order. Using realistic parameters for experimentally accessible heterostructures, we demonstrate that these signatures are experimentally measurable, establishing Zeeman quantum geometry as a powerful framework for characterizing unconventional magnetic insulators via gyrotropic transport responses.
\end{abstract}

\maketitle

\section{Introduction}
Unconventional magnets, characterized by momentum-dependent and symmetry-enforced spin-split band structures, have emerged as a paradigm-shifting class of magnetic materials. A prominent example is the altermagnetic family, in which adjacent collinear magnetic moments alternate in sign as a consequence of a unique nonrelativistic spin-group symmetry~\cite{Smejkal2022, Ahn2023, Mazin2024, Cheong2025}. Unlike conventional collinear antiferromagnets, altermagnets explicitly break time-reversal symmetry (TRS), generating strong momentum-space ($\mathbf{k}$-space) spin polarization and band splitting while maintaining zero net magnetization~\cite{Hayami2019, Jungwirth2022}. Depending on the parity of the magnetic order parameter, these systems effectively realize either even-parity magnetism, also called altermagnets (e.g., $d$-wave), which break TRS while preserving inversion symmetry, or odd-parity magnetism (e.g., $p$-wave), which preserves TRS but breaks inversion~\cite{Libor_2023,Linder_2024_PRL}. A defining feature of such unconventional magnetic orders is that the momentum-dependent spin splitting vanishes at symmetry-enforced nodal points in the Brillouin zone, with the nodal structure determined by the underlying magnetic form factor.

Transport phenomena in these systems are intimately connected to the quantum geometric tensor (QGT), which encodes the geometry of Bloch states in Hilbert space. The QGT has been identified as the microscopic origin of a variety of unconventional Hall and longitudinal responses in altermagnets, including higher-order anomalous Hall effects governed jointly by the Berry curvature and the quantum metric tensor~\cite{Fang_2024, Chen_2025,ezawa2025,chakraborti2025,Ezawa2025c,Song_2025}. Extending this notion, the recently introduced Zeeman quantum geometric tensor (ZQGT) incorporates the combined effects of momentum translations and spin rotations~\cite{Xiang_2025_PRL,chakraborti2026PST,chakraborti2026spin}. While previous studies have primarily focused on electric-field-driven transport governed by the conventional QGT~\cite{Fang_2024,Chen_2025,ezawa2025}, the role of the ZQGT in magnetic-field-driven transport and its application to unconventional magnetic insulators remain largely unexplored.

Despite rapid theoretical progress, the unambiguous identification of unconventional magnetic phases remains experimentally challenging. For instance, while $\mathrm{RuO_2}$ was initially proposed as a prototypical altermagnet, subsequent investigations have complicated this classification, underscoring the need for cleaner and more controllable experimental platforms capable of resolving $\mathbf{k}$-space spin textures~\cite{Feng2022,Sinova2020,Gonzalez2021}. In this context, topological insulators provide an ideal setting owing to their spin–momentum–locked surface states, which are inherently sensitive to symmetry breaking. In conventional magnetic topological insulators (MTIs), such as Cr- or Mn-doped Bi$_2$(Se,Te)$_3$, an exchange gap is opened through the coupling between surface Dirac electrons and localized magnetic moments~\cite{Chang2013,Tokura2019}. The successful realization and manipulation of these phases, manifested for example in the quantum anomalous Hall effect~\cite{Chang2013}, axion-insulator states~\cite{Mogi2017, Liu2020}, and half-quantized Hall conductance~\cite{Tokura2019, Deng2020}; naturally raise a key question: how does the transport response of the surface states of a topological insulator change when it is subjected not to a uniform ferromagnetic exchange field, but instead to a momentum-dependent, sign-alternating exchange field generated by an unconventional magnet?

Here, we address this question by developing a unified framework for the intrinsic gyrotropic magnetic (IGM) response of a three-dimensional MTI–unconventional magnetic insulator (UMI) bilayer heterostructure. Exploiting the extreme sensitivity of MTI surface states to symmetry-breaking perturbations, we employ the quantum kinetic theory incorporating the ZQGT to compute the linear gyrotropic response for magnetic order parameters with $p$-, $d$-, $f$-, and $g$-wave symmetries~\cite{Xiang_2025_PRL}. We demonstrate that both the angular dependence and the sign structure of the IGM response are dictated entirely by the parity and rotational symmetry of the magnetic form factor. In particular, the transverse displacement IGM conductivity exhibits exclusively even-fold angular harmonics reflecting the underlying unconventional magnetic character, whereas the longitudinal displacement IGM conductivity encodes the even or odd-parity nature of unconventional magnetic order. These oscillations act as symmetry-enforced fingerprints, establishing the ZQGT-mediated IGM response as a precise, non-invasive diagnostic for distinguishing unconventional magnetic orders.

The remainder of the paper is organized as follows. In \sect{sec:IGM}, we introduce the generalized quantum geometry arising from the combined momentum translation and spin rotation of Bloch states, and derive the expressions for the linear IGM conductivities induced by the Zeeman quantum geometry. In \sect{sec:model}, we present the effective model Hamiltonian describing the surface states of the MTI–UMI bilayer heterostructure. Our main results for the IGM response associated with various unconventional magnetic textures are discussed in \sect{sec:results}. Finally, in \sect{sec:summary}, we conclude with a summary of our findings, a discussion of experimental feasibility, and an outlook on future directions.

 \section{Intrinsic Gyrotropic Magnetic Response driven by Generalized Quantum Geometry}
 \label{sec:IGM}
The quantum geometric tensor is defined from the quantum distance between Bloch states $\ket{u^{\xi}_{m\mathbf{k}}}$ differing by infinitesimal momentum shifts, with $m$, $\mathbf{k}$, and $\xi$ denoting the band, momentum, and spin indices, respectively. A generalized formulation incorporating both momentum translations and spin rotations allows the quantum distance to be written as~\cite{Xiang_2025_PRL}
\begin{align}
ds^2 &= \left\lVert {\mathscr R}_{d\theta} {\mathscr T}_{d\mathbf{k}} \ket{u_{m\mathbf{k}}^\xi} - \ket{u_{m\mathbf{k}}^\xi} \right\rVert^2 \nonumber \\
&= \sum_{p \neq m} {\mathscr G}_{mp}^{ab} \, dk_a dk_b 
+ \frac{1}{4} \sum_{m} {\mathcal S}_{pm}^{ab} \, d\theta_a d\theta_b \nonumber \\ 
&\quad + \frac{1}{2} \sum_{p \neq m} \left({\mathscr Z}_{mp}^{ba} + {\mathscr Z}_{pm}^{ba}\right) d\theta_a dk_b,
\label{eq:qgt}
\end{align}
where $a,b$ denote spatial indices. Here, ${\mathscr T}_{d\mathbf{k}} = e^{-i d\mathbf{k}\cdot\hat{\mathbf{r}}}$ and ${\mathscr R}_{d\theta} = e^{-i d\theta\cdot\hat{\boldsymbol{\sigma}}/2}$ generate infinitesimal momentum translation by $d\mathbf{k}$ and spin rotation by ${d\theta}$, respectively. The conventional QGT, ${\mathscr G}_{mp}^{ab}$, arises from momentum translations, while the spin-rotation QGT, ${\mathcal S}_{pm}^{ab}$, stems from spin rotations. The Zeeman QGT, ${\mathscr Z}_{mp}^{ab} = r_{mp}^a\sigma_{pm}^b = \mathcal{Q}_{mp}^{ab} - \frac{i}{2}\mathcal{Z}_{mp}^{ab}$, couples both processes. The Zeeman quantum metric (ZQM) ($\mathcal{Q}_{mp}^{ab}$) and the Zeeman Berry curvature (ZBC) ($\mathcal{Z}_{mp}^{ab}$) are given by~\cite{chakraborti2026}
\begin{align}
\mathcal{Q}_{mp}^{ab} &= \tfrac{1}{2}\left(r_{mp}^a\sigma_{pm}^b + r_{pm}^a\sigma_{mp}^b\right), \nonumber\\
\mathcal{Z}_{mp}^{ab} &= i\left(r_{mp}^a\sigma_{pm}^b - r_{pm}^a\sigma_{mp}^b\right).
\label{eq:ZQGT}
\end{align}
The ZBC (ZQM) is even (odd) under time-reversal symmetry, while both are odd under inversion. Moreover, ZQM and ZBC have both symmetric and antisymmetric components in general.

\begin{figure}[!b]
\centering
\includegraphics[width=\columnwidth]{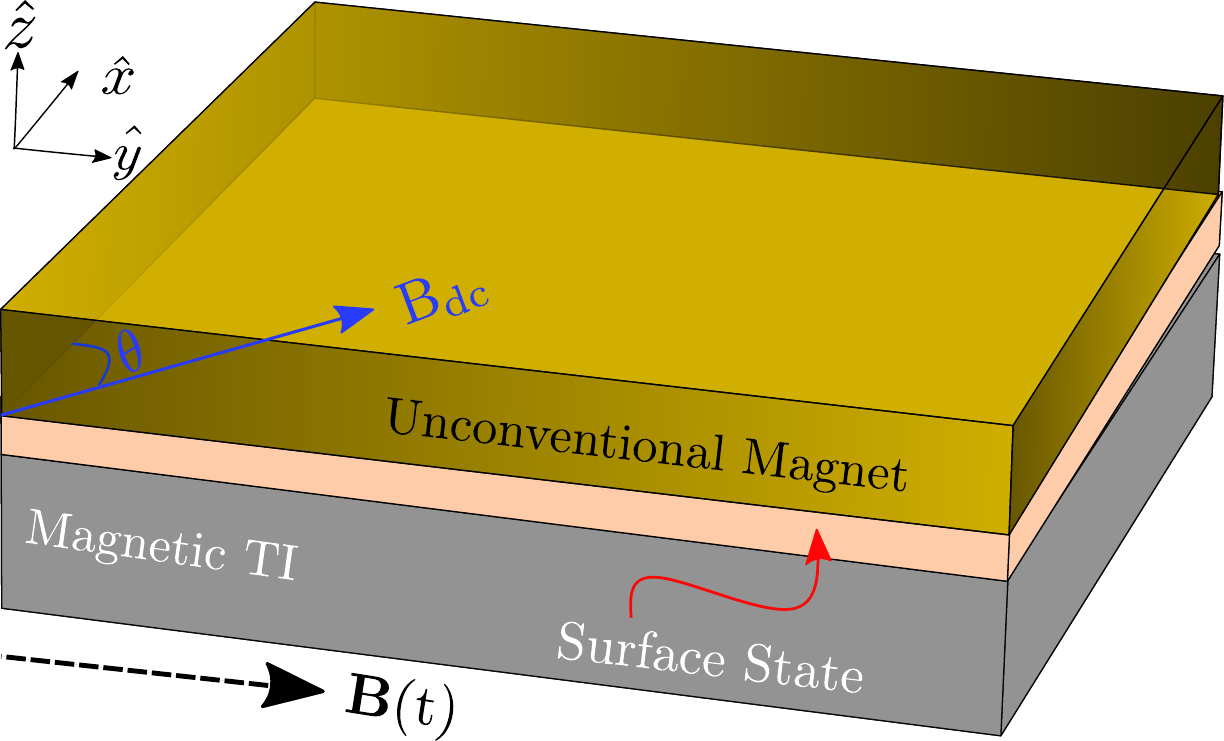}
\caption{\textbf{Schematic of a magnetic topological insulator (MTI)–unconventional magnetic insulator (UMI) bilayer device.} A heterostructure composed of a ferromagnetic topological insulator (e.g. Cr-doped Bi$_2$Se$_3$) interfaced with an insulating unconventional magnet is shown. An in-plane dc magnetic field ($\mathbf{B}_{\rm dc}$) shifts the surface Dirac cone of the MTI from the $\Gamma$-point to a finite in-plane momentum. The proximity-induced, momentum-dependent magnetic exchange field generated by the unconventional magnet opens a Dirac mass gap whose sign varies in momentum space. The intrinsic gyrotropic magnetic (IGM) response of the device is then probed by applying a weak, time-dependent ac magnetic field $\mathbf{B}(t)$.
}
\label{fig:Schematic}
\end{figure}

\medskip
\noindent
A weak oscillating ac magnetic field $\mathbf{B}(t) = \mathbf{B}_{\rm ac} \cos(\omega t)$ with frequency $\omega$ gives rise to two types of IGM conductivities: the conduction response $\sigma^{C}_{ab}$ and the displacement response $\sigma^{D}_{ab}$. These are governed by the ZBC and ZQM, and up to linear order in $\omega$ can be written as~\cite{Xiang_2025_PRL}:
\begin{equation}
\sigma^C_{ab} = \sum_{mp} \!\int_{\mathbf{k}} f_m \mathcal{Z}^{ab}_{mp}, \qquad
\sigma^D_{ab} = \sum_{mp} \!\int_{\mathbf{k}} f_m \frac{2\hbar\omega}{\epsilon_{pm}} \mathcal{Q}^{ab}_{mp},
\label{eq:Current}
\end{equation}
where $f_m=\frac{1}{e^{(\epsilon_m-\mu)/(k_B T)}+1}$ is the Fermi-Dirac distribution, $k_B$ is the Boltzmann constant, $\mu$ is the chemical potential and $\epsilon_{pm} = (\epsilon_p - \epsilon_m)$ is the interband energy difference. We consider frequencies below the interband absorption threshold ($\hbar \omega \ll \epsilon_{pm}$). It is clear from the expressions that the conduction IGM conductivity is a Fermi-surface quantity, while the displacement IGM conductivity is a Fermi-sea quantity. Both originate purely from band geometric quantities and are independent of the relaxation time, confirming their intrinsic nature.

\section{Model}
\label{sec:model}

To investigate the interplay between topological insulator surface states and unconventional magnetism on the IGM conductivities, we consider an MTI–unconventional magnet bilayer heterostructure, schematically shown in \fig{fig:Schematic}. The system consists of a ferromagnetic topological insulator, such as Cr-doped Bi$_2$Se$_3$~\cite{Tokura2019}, interfaced with an insulating unconventional magnetic layer. In this heterostructure, the exchange interaction between the localized magnetic moments of the dopants and the spin of the surface electrons gives rise to an effective Hamiltonian for the MTI surface states. When this interaction is confined to the in-plane direction by an externally applied dc magnetic field $\mathbf{B}_{\rm dc}$ acting on the magnetic dopants, the massless surface Dirac cone at the $\Gamma$ point is shifted to a finite momentum in the surface Brillouin zone. The in-plane magnetic field is applied at an azimuthal angle $\theta$ measured from the $x$ axis~\cite{Chen_2025}. 

In addition, the proximity effect induced by the insulating unconventional magnet generates a momentum-dependent Dirac mass for the MTI surface states~\cite{Chen_2025}. Implementing open boundary conditions along the $z$ axis, the resulting effective Hamiltonian describing the surface states of the MTI in the bilayer heterostructure can be written as~\cite{Chen_2025}:
\begin{equation}
H_0 = v_F \mathbf{k'} \cdot \sigma + g_{\mathbf{k}}\sigma_z,
\label{eq:H0}
\end{equation}
where $\mathbf{k'} = (\mathbf{k} - \mathbf{k}^{0})$, and $\sigma$ denotes the Pauli matrices acting on the spin degrees of freedom. Here, $v_F$ is the Fermi velocity, while $\mathbf{k}^{0}$ represents the momentum shift of the Dirac cone induced by the Zeeman energy scale $\Delta_{ex}$ generated by the in-plane magnetic field $\mathbf{B}_{\rm dc}$ acting on the magnetic dopants. The momentum shift is explicitly given by $\mathbf{k}^{0} = \frac{\Delta_{ex}}{v_F}(\cos\theta, \sin\theta)$, where $\theta$ is the azimuthal angle of the applied in-plane magnetic field. 

The term $g_{\mathbf{k}}$ encodes the form factor of the unconventional magnetic order, which imprints a momentum-dependent and sign-alternating Dirac mass onto the otherwise massless surface Dirac cone of the topological insulator. In the absence of magnetic interactions ($\Delta_{ex} = g_{\mathbf{k}} = 0$), the surface states of the topological insulator preserve time-reversal symmetry while breaking inversion symmetry, giving rise to a massless Dirac cone centered at the $\Gamma$ point. The Zeeman field $\Delta_{ex}$ explicitly breaks TRS and shifts the Dirac cone to the finite momentum $(k_x^0, k_y^0)$. The proximity-induced term $g_{\mathbf{k}}\sigma_z$ further breaks spin-rotation symmetry and introduces a momentum-dependent Dirac mass at the shifted Dirac cone, thereby encoding the spin texture of the underlying unconventional magnet into the electronic structure of the MTI surface states.

\section{Results}
\label{sec:results}

In this section, we analyze the IGM responses of the device shown in \fig{fig:Schematic} for different unconventional magnetic orders, including $p$-, $d$-, $f$-, $g$-, and $i$-wave symmetries, and highlight their distinct symmetry- and parity-dependent signatures. We demonstrate that the displacement and conduction components of the IGM conductivity provide complementary information, as discussed in detail below.

\subsection{IGM response of the MTI surface states}

\begin{figure}[!b]
\centering
\includegraphics[width=\columnwidth]{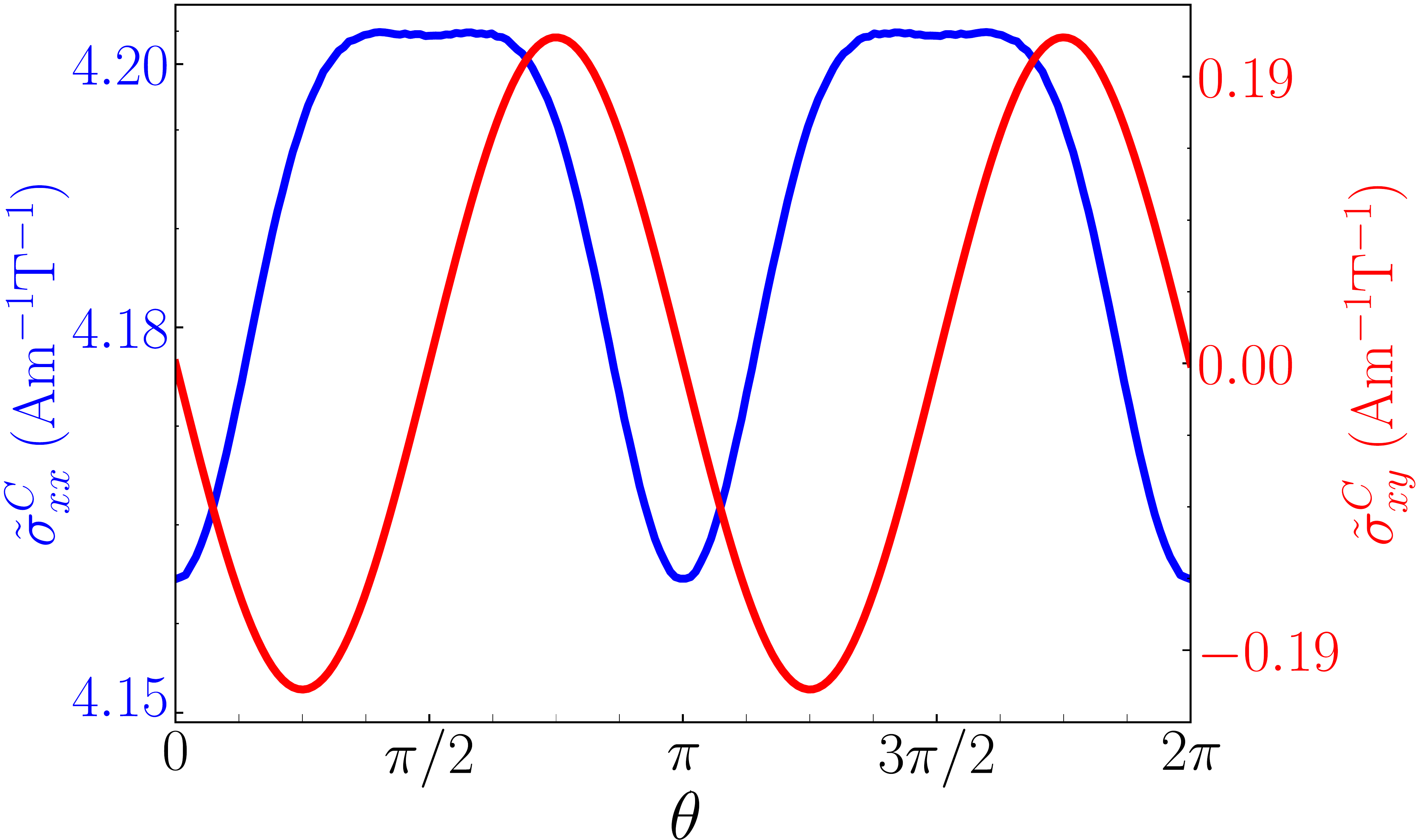}
\caption{\textbf{Conduction IGM conductivity of massless Dirac surface states of the MTI:} $\sigma_{xx}^{C}$ and $\sigma_{xy}^{C}$ components of the IGM conductivities are depicted for the massless Dirac limit $(g_{\mathbf{k}}=0)$. The parameters used are $v_F = 1~\text{eV}$, $\Delta_{ex} = 0.1~\text{eV}$, $T = 10~\text{K}$, $\mu = 0~\text{eV}$ and  $\tilde\sigma_{ab}^{C} = -g \mu_B \sigma^C_{ab}$.
}
\label{fig:massless_Dirac}
\end{figure}

In the absence of proximity coupling to an unconventional magnet, the effective surface Hamiltonian of the MTI hosts a massless Dirac cone shifted to a finite momentum determined solely by the applied in-plane dc magnetic field. In this limit, the Zeeman quantum metric vanishes identically, and as a result, the displacement current contribution to the IGM response is absent. In contrast, the Zeeman Berry curvature becomes finite and is given by
$Z^{xy}_{\pm \mp} = \mp \frac{v_F^2 k'_x k'_y}{2k'^{3}} = - Z^{yx}_{\pm \mp}$ and 
$Z^{xx}_{\pm \mp} = \mp \frac{v_F^2 {k'_y}^2}{2{k'}^{3}} = Z^{yy}_{\pm \mp}$,
where $k'$ denotes the magnitude of the in-plane wave vector $\mathbf{k'}$. The variation of the corresponding conduction IGM conductivities with the angle of the applied dc magnetic field are shown in \fig{fig:massless_Dirac}. As evident from the figure, the longitudinal component $\sigma_{xx}^{C}$ remains positive for all values of the angle $\theta$ and exhibits a characteristic $\sin^2(\theta)$ modulation. In contrast, the transverse component $\sigma_{xy}^{C}$ displays a $\sin(2\theta)$ angular dependence.

\begin{figure*}[ht!]
\centering
\includegraphics[width=\textwidth]{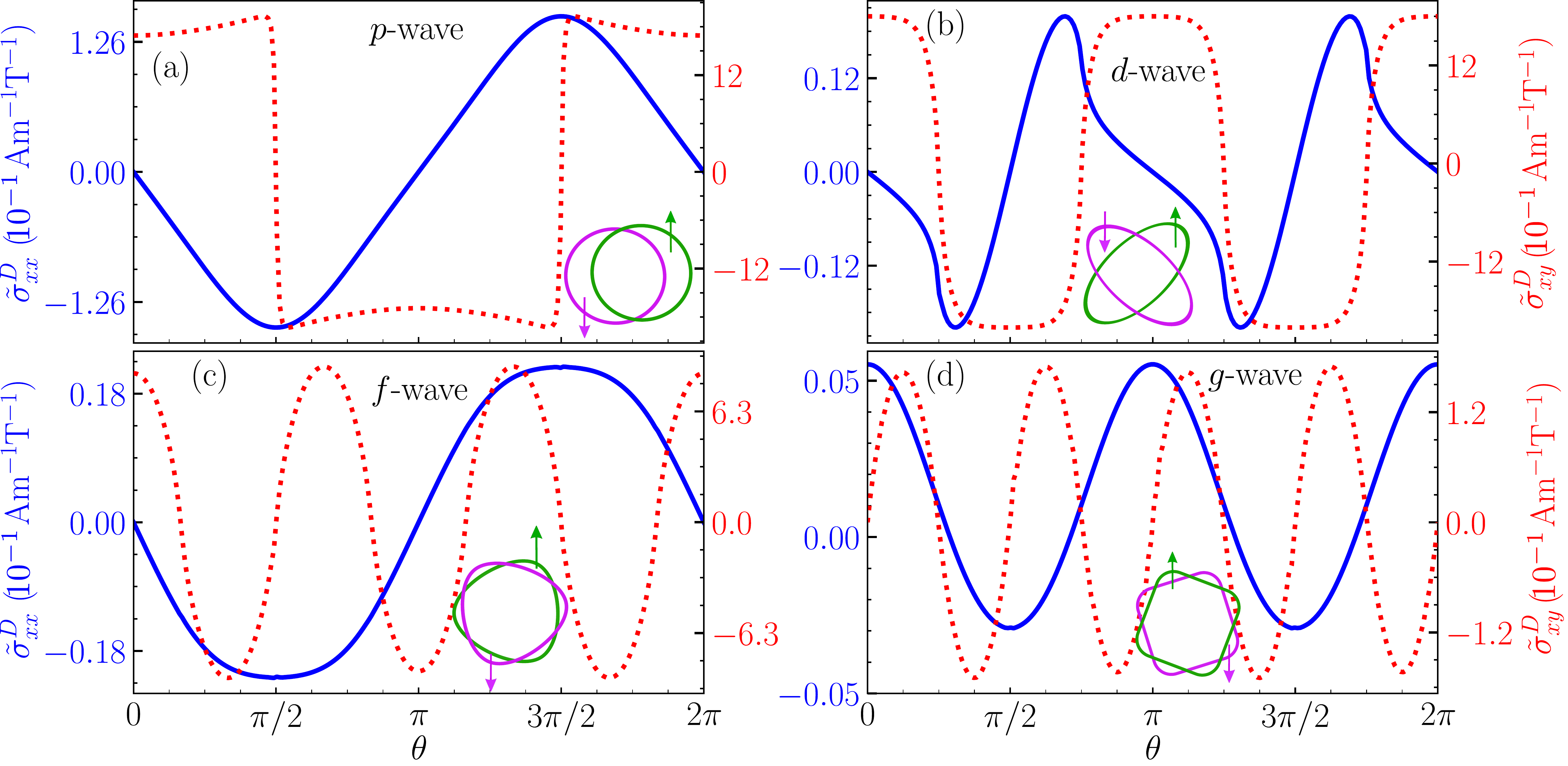}
\caption{\textbf{Distinguishing the parity and the structure of order parameter of unconventional magnets from displacement IGM conductivity:} (a) Longitudinal displacement component $\sigma^{D}_{xx}$(blue curve) for different unconventional magnetic textures ($p$-, $d$-, $f$-, and $g$-wave), showing symmetry-dependent angular modulations that distinguish the odd- and even-parity classes. (b) Transverse displacement component $\sigma^{D}_{xy}$(red curve) for the same textures, exhibiting the characteristic $2n$-fold sign-changing pattern set by the angular order of the exchange field. The parameters used are $v_F = 1~\text{eV}$, $J_0 = 0.3~\text{eV}$, $\Delta_{ex} = 0.1~\text{eV}$, $T = 10~\text{K}$, $\mu = 0~\text{eV}$, $\omega = 10^{13} \text{Hz}$ and $\tilde\sigma_{ab}^{D} = -g \mu_B \sigma^D_{ab}$.}
\label{fig:IGM_currents}
\end{figure*}

\subsection{IGM response at charge neutrality ($\mu = 0$)} due to proximity to unconventional magnets

We now focus on the role of the proximity effect induced by different types of unconventional magnetic orders (e.g. $p$-, $d$-, $f$-, $g$-, and $i$-wave) on the intrinsic gyrotropic magnetic currents at the charge neutrality point ($\mu = 0$). We first emphasize that the conduction IGM conductivities ($\sigma_{xy}^{C}$ and $\sigma_{xx}^{C}$) remain qualitatively identical to those of the massless Dirac cone discussed in \fig{fig:massless_Dirac} for all unconventional magnetic textures considered here. This insensitivity arises because the conduction current originates from the TRS-even Zeeman Berry curvature, which contains two contributions: one proportional to the unconventional magnetic form factor $g_{\mathbf{k}}$ and another that is independent of $g_{\mathbf{k}}$. In the presence of proximity-induced magnetism, the ratio between the $g_{\mathbf{k}}$-dependent and $g_{\mathbf{k}}$-independent terms scales as $\left( \frac{J_0}{v_F} \right)^2 \ll 1$, where $J_0$ denotes the strength of the unconventional magnetic order. Consequently, the $g_{\mathbf{k}}$-independent contribution dominates, rendering the ZBC—and hence the conduction IGM response—largely insensitive to the detailed momentum-space spin texture of the unconventional magnet. Note that the applied ac magnetic field is assumed to be sufficiently weak, such that the corresponding Zeeman energy ($\sim 1.2,\mu\text{eV}$) is several orders of magnitude smaller than the exchange energy $J_0$ ($\sim 0.3,\text{eV}$). Consequently, the ac field acts only as a linear-response probe of the electronic states, allowing the proximity-induced magnetic texture to be treated as static.

In sharp contrast, the displacement current originates from the TRS-odd Zeeman quantum metric and is directly proportional to the strength of the unconventional magnetic order $J_0$. For the Hamiltonian in Eq.~(\ref{eq:H0}), the longitudinal and transverse components of the ZQM are given by $\mathcal{Q}_{ii}=\frac{k'_y\partial_{i}g_{\mathbf{k}}}{2\beta_{\mathbf{k}}^3}$ and $\mathcal{Q}_{ij}=\frac{k'_x\partial_{i}g_{\mathbf{k}}-g_{\mathbf{k}}}{2\beta_{\mathbf{k}}^3} \approx -\frac{g_{\mathbf{k}}}{2\beta_{\mathbf{k}}^3}$, respectively, where $\beta_{\mathbf{k}}= \sqrt{{k'_x}^2 + {k'_y}^2 + g_k^2}$. These expressions make it evident that $\mathcal{Q}_{ij}$ is directly proportional to $g_{\mathbf{k}}$, implying that the momentum-dependent spin texture encoded in the Dirac mass of the unconventional magnet is faithfully reflected in the off-diagonal components of the ZQM. Since the transverse displacement current depends explicitly on $\mathcal{Q}_{ij}$, it therefore provides a highly sensitive probe of the underlying unconventional magnetic texture.

By contrast, the longitudinal displacement current depends on $\mathcal{Q}_{ii}$, which is proportional to $\partial_{i} g_{\mathbf{k}}$, and hence captures the parity (even or odd) of the magnetic form factor. In light of these considerations, we focus exclusively on the displacement IGM conductivities, which offer a direct experimental route to probing the local $\mathbf{k}$-space spin textures of unconventional magnets, complementary to recently proposed anomalous Hall probes~\cite{Chen_2025}.

\subsubsection{$p$-wave unconventional magnet}

A $p_x$-wave unconventional magnet is characterized by an odd-parity exchange texture 
$g^{p}_{\mathbf{k}} = J_0 k_x$, which satisfies $g^{p}(-k_x) = -g^{p}(k_x)$~\cite{ezawa2025,Ezawa2025c,Brekke2024}. 
This form ensures that the induced proximity term remains invariant under the mirror operation 
$M_x: (k_x, k_y) \rightarrow (-k_x, k_y)$. 
It explicitly breaks full rotational symmetry while preserving the combined $C_{2z}\mathcal{T}$ symmetry, where $C_{2z}$ denotes a twofold rotation about the $z$ axis~\cite{hellenes2024,jungwirth2024b}. 
Importantly, the $p_x$-wave form factor vanishes along the line $k_x = 0$, giving rise to two nodes in momentum space where opposite spin sectors touch.

The longitudinal and transverse components of the displacement IGM conductivities, 
$\sigma_{xx}^{D}$ and $\sigma_{xy}^{D}$, as functions of the polar angle $\theta$ for the MTI surface states described by Eq.~(\ref{eq:H0}), 
are shown in Fig.~\ref{fig:IGM_currents}(a). 
Strikingly, as $\theta$ increases, the transverse component $\sigma_{xy}^{D}$ exhibits two sign reversals near $\theta = \pi/2$ (positive to negative) and $3\pi/2$ (negative to positive) within the interval $\theta \in [0, 2\pi]$. The sign of $\sigma_{xy}^{D}$ is governed by the odd-parity exchange texture $g_{\mathbf{k}}^{p}$, whose sign follows that of $k_x$—positive in the first and fourth quadrants and negative in the second and third quadrants. As $\theta$ increases, the massive Dirac cone shifts toward the $-k_x$ direction, reaching $(0,\Delta_{ex}/v_F)$ at $\theta=\pi/2$, where $g_{\mathbf{k}}^{p}=0$ and hence $\sigma_{xy}^{D}=0$. Beyond this angle, the cone enters the second quadrant, reversing the sign of $g_{\mathbf{k}}^{p}$ and consequently that of $\sigma_{xy}^{D}$. This behavior persists until $\theta=3\pi/2$, where the Dirac cone again crosses $k_x=0$ before entering the fourth quadrant and restoring a positive sign. The periodicity of $\sigma_{xy}^{D}$ follows from the $C_{2z}\mathcal{T}$ symmetry of the $p_x$-wave magnet, yielding $\sigma_{xy}^{D}(\theta)=\sigma_{xy}^{D}(\theta+\pi)$. These alternating sign changes provide a direct diagnostic of the nodal structure and momentum-dependent spin texture associated with $p$-wave magnetism.

The longitudinal component $\sigma_{xx}^{D}$ exhibits two sign reversals at $\theta = 0$ and $\pi$. 
For a $p$-wave magnet, $Q_{xx} = \frac{J_{0}\,k'_{y}}{\beta^{3}}$, which necessarily vanishes at two points in momentum space. The sign of $\sigma_{xx}^{D}$ is therefore controlled by the sign of $k'_{y}$, which changes twice as the Dirac cone traverses the Brillouin zone. This behavior directly reflects the odd-parity character of the $p$-wave magnetic order. We note that for the $p_y$-wave case, $\sigma_{xx}^{D}$ vanishes identically, while $\sigma_{yy}^{D}$ exhibits two nodes. The transverse component $\sigma_{xy}^{D}$ also displays two nodes, whose angular positions are shifted relative to the $p_x$-wave case due to the rotated exchange texture.

\subsubsection{$d$-wave altermagnet}

For a $d_{x^2 - y^2}$-wave unconventional magnet (also referred to as a $d_{x^2 - y^2}$-wave altermagnet) characterized by the even-parity exchange texture $g_{\mathbf{k}}^d=J_0(k_y^2-k_x^2)$~\cite{smejkal2022prx,roig2024,ezawa2025,Ezawa2025c,Smejkal2022Nature,Ghorashi2024PRL}, the proximity-induced term preserves a combined $C_{4z}\mathcal{T}$ symmetry and produces a spin-polarized Fermi surface with diagonal degeneracies along $k_x = \pm k_y$. The longitudinal ($\sigma_{xx}^{D}$) and transverse ($\sigma_{xy}^{D}$) components of the displacement IGM conductivities as functions of the polar angle $\theta$ are shown in Fig.~\ref{fig:IGM_currents}(b). Unlike the $p_x$-wave case, the transverse component $\sigma_{xy}^{D}$ undergoes four sign reversals and vanishes at $\theta = n\pi/4$ ($n = 1, 3, 5, 7$) within the interval $\theta \in [0, 2\pi]$. This behavior originates from the momentum-dependent Dirac mass $J_0 (k_x^2 - k_y^2)$, which induces opposite mass inversions in adjacent angular sectors. Consequently, a sign change occurs whenever the condition $k_y > k_x$ is satisfied, which happens once in each quadrant. The vanishing of $\sigma_{xy}^{D}$ at $\theta = n\pi/4$ reflects the four nodes of the $d_{x^2-y^2}$-wave form factor. At these angles, the Dirac points are located at $\frac{\Delta_{\rm ex}}{v_F}\left(\frac{1}{\sqrt{2}}, \pm \frac{1}{\sqrt{2}}\right)$ and $\frac{\Delta_{\rm ex}}{v_F}\left(\pm \frac{1}{\sqrt{2}}, \frac{1}{\sqrt{2}}\right)$, where $g_{\mathbf{k}}^d$ vanishes, yielding $\sigma_{xy}^{D} = 0$. 

The longitudinal component $\sigma_{xx}^{D}$ also exhibits four sign reversals over $\theta \in [0, 2\pi]$, as $Q_{xx}$ scales as $\frac{J_{0} k_x {k'}_{y}}{\beta^{3}_{\mathbf{k}}}$ and necessarily vanishes at four distinct momenta. This fourfold structure directly reflects the even-parity nature of the underlying $d$-wave altermagnetic order. For completeness, we note that in the related $d_{xy}$-wave altermagnet, $\sigma_{xy}^D$ and $\sigma_{xx}^D$ also exhibit four and two sign changes, respectively, but with rotated nodal positions: the nodes of $\sigma_{xy}^D$ shift to $\theta = n\pi/2$ ($n = 0, \dots, 3$), while those of $\sigma_{xx}^D$ appear at $\theta = n\pi/4$ ($n = 1, 3, 5, 7$).

\begin{table*}[!htb]
\centering
\renewcommand{\arraystretch}{1.5}
\setlength{\tabcolsep}{7.0pt}

\begin{tabular}{|c|c|c|c|c|}
\hline
Different magnets
& Symmetry 
&$\sigma_{xy}^D$ &$\sigma_{xx}^D$ & Possible Materials\\ 
\hline
$p_x$-wave & $C_{2z}\mathcal{T}, \mathcal{T}$ & $\sigma^D_{xy} = - \sigma^D_{yx}(\text{two nodes})$ & $\sigma^D_{xx} (\text{two nodes})$ & CeNiAsO~\cite{hellenes2024}, Ni$\text{I}_2$~\cite{Song2025}\\
\hline
$d_{x^2-y^2}$-wave & $C_{4z}\mathcal{T}, \mathcal{P}$ & $\sigma^D_{xy} = - \sigma^D_{yx}(\text{four nodes})$ & $\sigma^D_{xx} = \sigma^D_{yy}(\text{four nodes})$ & Ru$\text{O}_2$~\cite{Ahn2023}, $\text{Mn}_5\text{Si}_3$~\cite{Leiviska2024}\\
\hline
$f$-wave & $C_{6z}\mathcal{T},  \mathcal{T}$ & $\sigma^D_{xy} = -\sigma^D_{yx}(\text{six nodes})$ & $\sigma^D_{xx} \neq \sigma^D_{yy} (\text{two nodes})$ & $\text{Ba}_3\text{Mn}\text{Nb}_2 \text{O}_9$~\cite{Lee2014}\\
\hline
$g$-wave & $C_{8z}\mathcal{T}, \mathcal{P}$ & $\sigma^D_{xy} = -\sigma^D_{yx}(\text{eight nodes})$ & $\sigma^D_{xx} = \sigma^D_{yy} (\text{four nodes})$ & MnTe~\cite{Lee2024}, CrSb~\cite{Yang2025,Ding2024,Li2025,wan2011,Reimers2024}\\
\hline
$i$-wave & $C_{12z}\mathcal{T}, \mathcal{P}$ & $\sigma^D_{xy} = -\sigma^D_{yx}(\text{twelve nodes})$ & $\sigma^D_{xx} = \sigma^D_{yy} (\text{four nodes})$ &  \shortstack{Twisted Magnetic \\ Van der Waals Bilayers}~\cite{Liu2024_prl, korrapati2025}\\
\hline
\end{tabular}
\caption{Summary of the angular symmetries of unconventional magnetic orders and the corresponding nonvanishing components of the displacement-type IGM conductivity $\sigma^D_{\alpha\beta}$. The number and arrangement of angular nodes directly encode the underlying $C_{nz}\mathcal{T}$ rotational symmetry and the parity of the $p$-, $d$-, $f$-, $g$-, and $i$-wave unconventional magnetic orders.}
\label{tab:Summery}
\end{table*}

\subsubsection{$f$-wave unconventional magnet}

For $f$-wave unconventional magnets characterized by an odd-parity exchange texture $g_{\mathbf{k}}^f = J_0\,k_x(k_x^2 - 3k_y^2)$~\cite{uchino2025, ezawa2025,Ezawa2025c}, the exchange field exhibits a cubic angular dependence with alternating sign lobes separated by $60^\circ$. This texture preserves a combined $C_{6z}\mathcal{T}$ symmetry and changes sign under inversion, with nodes along the lines $k_x = 0$ and $k_x = \pm \sqrt{3}\, k_y$. The corresponding displacement IGM conductivities, shown in Fig.~\ref{fig:IGM_currents}(c), directly encode this symmetry structure. The transverse component $\sigma_{xy}^{D}$ exhibits six sign reversals at $\theta = \pi/6 + n\pi/3$ ($n = 0, \dots, 5$) within $\theta \in [0, 2\pi)$. These zeros occur when the Dirac points align with the form-factor nodes—specifically at $\frac{\Delta_{\rm ex}}{v_F}(0,\pm1)$, $\frac{\Delta_{\rm ex}}{v_F}(\pm \sqrt{3},1)$, and $\frac{\Delta_{\rm ex}}{v_F}(1, \pm 1/\sqrt{3})$—causing the cubic term $(k_x^3 - 3k_x k_y^2)$ to vanish. As a result, $\sigma_{xy}^{D}$ alternates in sign with a periodicity of $\pi/3$, reflecting the alternating Dirac-mass structure characteristic of $f$- wave magnetism.

In contrast, the longitudinal component $\sigma_{xx}^{D}$ exhibits only two sign reversals at $\theta = 0$ and $\pi$, governed by the quantity $Q_{xx} \propto \frac{3J_{0}\,(k_x^{2} - k_y^{2})\,{\mathcal{K'}}_{y}}{\beta^{3}}$. This simpler angular dependence mirrors the qualitative behavior observed for the $p$-wave case and serves as a distinct signature of the odd-parity nature of the $f$-wave unconventional magnetic order.

\begin{figure*}[!ht]
\centering
\includegraphics[width=0.9\textwidth]{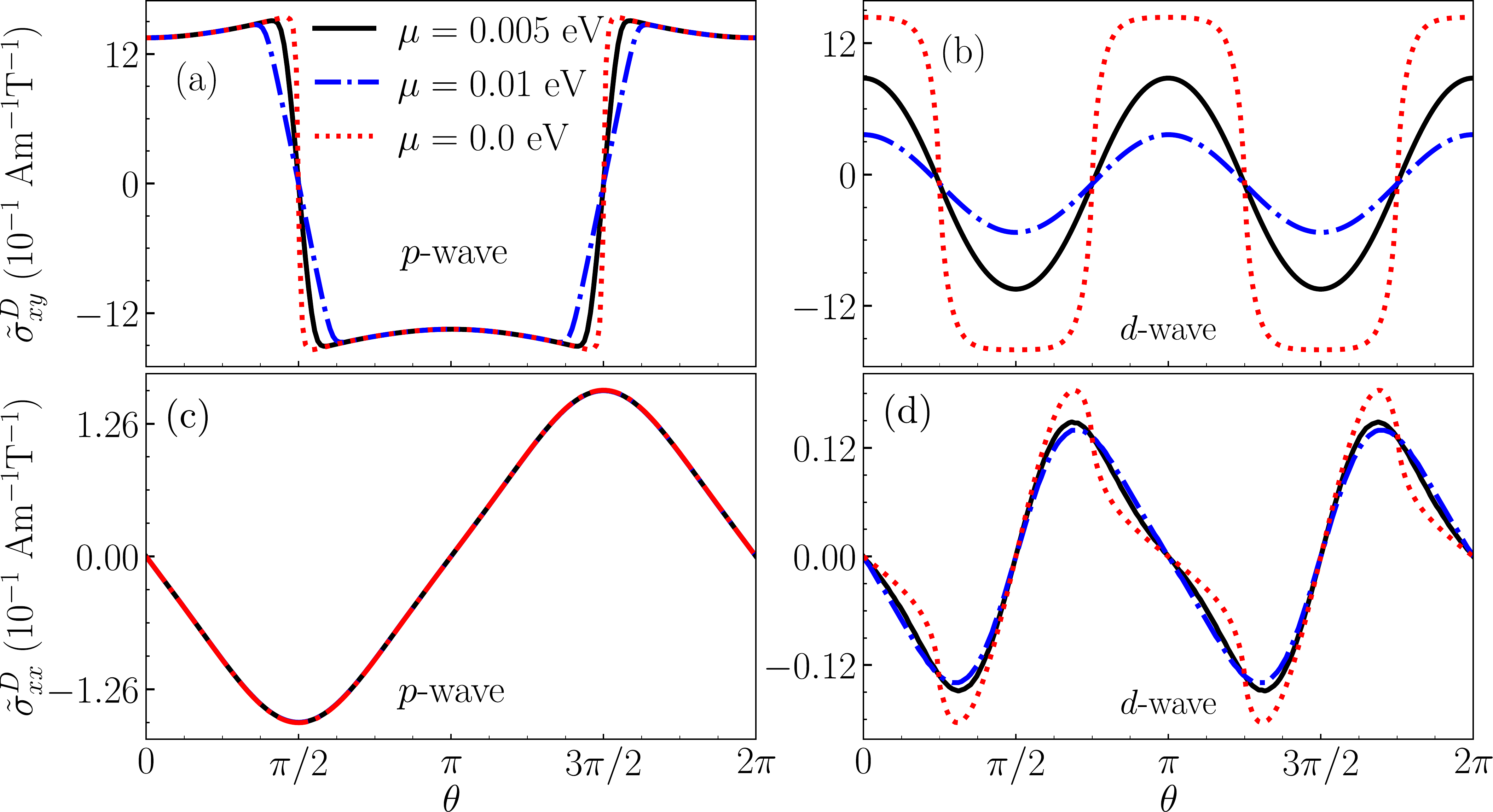}
\caption{\textbf{Chemical-potential dependence of the displacement IGM conductivity:} Variation of the (a,b) transverse displacement conductivity $\sigma^{D}_{xy}$ and (c,d) longitudinal displacement conductivity $\sigma^{D}_{xx}$ with chemical potential near the Dirac point for representative $p$- and $d$-wave unconventional magnetic orders, respectively.}
\label{fig:IGM_different_mu}
\end{figure*}

\subsubsection{$g$-wave altermagnet}

For a $g$-wave altermagnet characterized by the higher-order exchange texture $g_{\mathbf{k}}^g = 4J_0\,k_xk_y(k_x^2 - k_y^2)$~\cite{smejkal2022prx,roig2024,Karetta2025,ezawa2025,Ezawa2025c}, the exchange field exhibits a quartic angular dependence with alternating sign lobes separated by $45^\circ$, while preserving a combined $C_{8z}\mathcal{T}$ symmetry. The corresponding nodes lie along $k_x = 0$, $k_y = 0$, and the diagonals $k_x = \pm k_y$, partitioning momentum space into eight angular sectors. This structure is directly imprinted on the transverse displacement IGM conductivity $\sigma_{xy}^{D}$, shown in Fig.~\ref{fig:IGM_currents}(d), which exhibits eight sign reversals at $\theta = n\pi/4$ ($n = 0, \dots, 7$). These zeros coincide with Dirac points located at $\frac{\Delta_{\rm ex}}{v_F}(0,\pm 1)$, $\frac{\Delta_{\rm ex}}{v_F}(\pm 1,0)$, $\frac{\Delta_{\rm ex}}{v_F}\left(\frac{1}{\sqrt{2}}, \pm \frac{1}{\sqrt{2}}\right)$, and $\frac{\Delta_{\rm ex}}{v_F}\left(\pm \frac{1}{\sqrt{2}}, \frac{1}{\sqrt{2}}\right)$, where the form factor vanishes. Consequently, $\sigma_{xy}^{D}$ alternates in sign eight times over the full angular range, faithfully reflecting the alternating Dirac-mass pattern imposed by the $g$-wave altermagnetic symmetry.

The longitudinal component $\sigma_{xx}^{D}$ exhibits four sign reversals over the interval $\theta \in [0, 2\pi]$, governed by the quantity $Q_{xx}$, which scales as $\frac{3J_{0}\, k_y (3k_x^{2} - k_y^{2})\, k'_{y}}{\beta^{3}}$. This behavior, which mirrors the fourfold sign structure observed in the $d$-wave case, is dictated by the even-parity character of the underlying form factor and thus serves as a clear signature of the $g$-wave unconventional magnetic order.

\subsubsection{$i$-wave altermagnet}

For $i$-wave unconventional magnets characterized by an even-parity exchange texture
$g_{\mathbf{k}}^i = 2J_0\, k_x k_y (3k_x^2 - k_y^2)(k_x^2 - 3k_y^2)$, the exchange field exhibits a sixth-order angular dependence with twelve alternating sign lobes separated by $30^\circ$ in momentum space~\cite{ezawa2025,Ezawa2025c}. This texture is even under inversion and preserves a combined $C_{12z}\mathcal{T}$ symmetry. The corresponding form factor vanishes along multiple high-symmetry lines, including
$k_x=0$, $k_y=0$, $k_x=\pm\sqrt{3}\,k_y$, and $k_y=\pm\sqrt{3}\,k_x$. When the shifted Dirac points coincide with these nodal lines, the proximity-induced Dirac mass changes sign, leading to alternating contributions to the transverse displacement IGM response. As a result, $\sigma_{xy}^{D}$ exhibits a characteristic $\pi/6$ angular periodicity over the interval $\theta \in [0,2\pi)$, starting from $\theta = 0$, reflecting the higher-harmonic structure intrinsic to the $i$-wave symmetry (not shown here).

In contrast, the longitudinal displacement response $\sigma_{xx}^{D}$ displays four sign reversals over $\theta \in [0,2\pi)$, indicative of its even-parity origin and closely mirroring the behavior observed for the $d$- and $g$-wave cases. This clear separation between the angular harmonics of the transverse and longitudinal displacement currents establishes a distinct transport fingerprint for even-parity $i$-wave unconventional magnetic order.

\subsection{IGM response with varying chemical potential}
While the previous subsection focused on the case of $\mu = 0$, here we investigate the effect of a finite chemical potential, since in realistic experiments the chemical potential can be tuned by electrostatic gating or chemical doping. As representative examples, we show the variation of $\sigma_{ab}^{D}$ with $\mu$ for the $p$- and $d$-wave unconventional magnetic orders in the vicinity of the Dirac point in \fig{fig:IGM_different_mu}. We find that all the characteristic symmetry-derived features of $\sigma_{ab}^{D}$ discussed in the previous subsection remain robust for chemical potentials close to the Dirac point. Although varying $\mu$ slightly modifies the magnitude of the IGM response through changes in the electronic occupation, the characteristic angular modulation, node positions, and sign-reversal patterns remain essentially unchanged.

\section{Summary and Outlook}
\label{sec:summary}

In this work, we establish a direct correspondence between the parity and angular symmetry of unconventional magnetic orders and the induced intrinsic gyrotropic magnetic response of an MTI–UMI heterostructure, governed by the Zeeman quantum geometry. We identify a universal hierarchy in which each magnetic order—$p$-, $d$-, $f$-, $g$-, and $i$-wave—imprints a characteristic angular modulation onto the IGM transport signatures, as summarized in \tab{tab:Summery}. Specifically, the transverse displacement IGM conductivity $\sigma^D_{xy}$ exhibits even-fold ($2n$) angular harmonics, with $n = 1,2,3,4,6$ for $p$-, $d$-, $f$-, $g$-, and $i$-wave orders, respectively. These harmonics originate from the alternating Dirac mass structure associated with the unconventional magnetic exchange field and vanish upon integration over the Brillouin zone. The longitudinal displacement IGM conductivity $\sigma^D_{xx}$ further resolves the parity of the magnetic order, displaying four sign reversals for even-parity ($d$-, $g$-, and $i$-wave) magnetism and only two sign reversals for odd-parity ($p$- and $f$-wave) magnetism. In contrast, the conduction IGM components ($\sigma^{C}_{xx}$ and $\sigma^{C}_{xy}$) remain dominated by the Dirac dispersion and are largely insensitive to the symmetry of the magnetic order. While the present work focuses on the intrinsic linear IGM response, its nonlinear counterpart is also expected to provide complementary probes of unconventional altermagnetic textures through additional symmetry-sensitive higher-order response functions, making it an interesting direction for future investigation.

To assess experimental feasibility, we consider a representative bilayer heterostructure composed of Cr-doped Bi$_2$Se$_3$ interfaced with RuO$_2$, a proposed $d_{x^2-y^2}$-wave altermagnet. We focus on the regime $B_{\rm dc} \gg B_{\rm ac}$, where the in-plane dc magnetic field controls the orientation of the exchange field, while a weak oscillating ac magnetic field serves as the probe. For an in-plane dc field of $\sim 0.2$~T and 5\% Cr doping~\cite{Rui2010}, a Zeeman energy $\Delta_{ex} \approx 0.2$~eV is generated~\cite{Chang2013,Kandala2015}. Taking the strength of the $d_{x^2-y^2}$-wave order in RuO$_2$ to be $\sim 1$~eV and an in-plane ac magnetic field of amplitude $\sim 10$~mT at a frequency of $10^{13}$~Hz, both longitudinal and transverse displacement IGM currents can be driven. For a Bi$_2$Se$_3$ layer of thickness $\sim 100~\mu$m and resistance $\sim 10^2~\Omega$, we estimate a transverse displacement IGM voltage of $\sim 1.14$~mV and a corresponding longitudinal voltage of $\sim 0.11$~mV, placing both signals well within experimentally measurable ranges. Importantly, optical currents arising from interband transitions are suppressed in this regime since $\Delta_{ex} (0.2~\mathrm{eV}) \gg \hbar\omega (0.004~\mathrm{eV})$. Furthermore, by applying the probe magnetic field entirely within the plane of the heterostructure, the accompanying electric field generated via Faraday's law remains perpendicular to the probe field, thereby suppressing electric-field-induced quantum geometric responses, such as anomalous Hall currents. This measurement geometry helps isolate the Zeeman-driven intrinsic gyrotropic magnetic response.
Experimentally, the required in-plane magnetic-field angle can be accurately controlled using standard vector magnet systems or rotatable sample stages in low-temperature cryogenic transport measurements, making the proposed angular measurements experimentally feasible~\cite{Vincent_2024,Lewin_2024}. Although moderate impurity scattering may broaden the surface Dirac spectrum and smoothen the response, the characteristic angular harmonics and parity-dependent sign-reversal patterns are expected to remain robust, since transport is mediated by the spin-momentum-locked surface states of the topological insulator and these signatures are dictated by the symmetry of the proximity-induced exchange field and the associated Zeeman quantum geometry. These results, therefore, provide clear routes for detecting and distinguishing unconventional magnetic textures using magneto-optical and transport probes~\cite{Smejkal2022, Krempasky2024, Reimers2024,Wei2024, Smejkal2022b,Lee2016}. By explicitly linking the parity and angular structure of the exchange field to the harmonic content of the IGM response, our work establishes a robust electrical protocol for classifying unconventional magnetic orders. More broadly, this framework opens a pathway toward exploiting Zeeman quantum geometry to engineer novel functionalities in heterostructures combining topological insulators and unconventional magnetic materials.
By explicitly linking the parity and angular structure of the exchange field to the harmonic content of the IGM response, our work establishes a robust electrical protocol for classifying unconventional magnetic orders. More broadly, this framework opens a pathway toward exploiting Zeeman quantum geometry to engineer novel functionalities in heterostructures combining topological insulators and unconventional magnetic materials. 


\section{Acknowledgments}
N.~C. acknowledges the UGC, Government of India, for providing the SRF fellowship. S.~K.~G. and S.~N. acknowledge financial support from Anusandhan National Research Foundation (ANRF) erstwhile Science and Engineering Research Board (SERB), Government of India respectively via the Startup Research Grant: SRG/2023/000934 and the Prime Minister's Early Career Research Grant: ANRF/ECRG/2024/005947/PMS.

\section{Declarations}

\subsection{Ethical statement}
This work does not involve human participants, human data, or animals. Therefore, ethical approval and informed consent are not applicable.

\subsection{Competing Interests} 
The authors declare no competing interests.

\subsection{Funding}
S.~K.~G. and S.~N. acknowledge financial support from Anusandhan National Research Foundation (ANRF) erstwhile Science and Engineering Research Board (SERB), Government of India respectively via the Startup Research Grant: SRG/2023/000934 and the Prime Minister's Early Career Research Grant: ANRF/ECRG/2024/005947/PMS. 
\\
\subsection{Data Availability}
Data sharing not applicable to this article as no data sets were generated or analysed during the current study.

\subsection{Author contributions}
S.K.G. and S.N. have conceptualized and supervised the work. The calculations were performed by N.C. The paper was written by N.C. with inputs from S.K.G. and S.N. All authors have contributed to the discussions and analyses of the data and approved the final version.

\bibliography{SCDM}
\end{document}